\documentclass[a4paper,12pt,english]{article}
\usepackage[T1]{fontenc}
\usepackage{geometry}
\usepackage{babel,csquotes}
\usepackage{amssymb,amsmath}
\usepackage{tensor}
\usepackage{tikz}
\usepackage{hyperref}
\usepackage[
	sorting=none,
	maxnames=99,
	firstinits=true,
	doi=false,
	citestyle=numeric-comp,
	backend=biber
	]{biblatex}

\AtEveryBibitem{\clearfield{title}}
\renewbibmacro{in:}{}
\def\ddal{\mathop{\vrule height 7pt depth0.2pt
\hbox{\vrule height 0.5pt depth0.2pt width 6.2pt}\vrule height 7pt depth0.2pt width0.8pt
\kern-7.4pt\hbox{\vrule height 7pt depth-6.7pt width 7.pt}}}
\def\sdal{\mathop{\kern0.1pt\vrule height 4.9pt depth0.15pt
\hbox{\vrule height 0.3pt depth0.15pt width 4.6pt}\vrule height 4.9pt depth0.15pt width0.7pt
\kern-5.7pt\hbox{\vrule height 4.9pt depth-4.7pt width 5.3pt}}}
\def\ssdal{\mathop{\kern0.1pt\vrule height 3.8pt depth0.1pt width0.2pt
\hbox{\vrule height 0.3pt depth0.1pt width 3.6pt}\vrule height 3.8pt depth0.1pt width0.5pt
\kern-4.4pt\hbox{\vrule height 4pt depth-3.9pt width 4.2pt}}}
\def\dal{\mspace{1.5mu}\mathchoice{\ddal}{\ddal}{\sdal}{\ssdal}\mspace{1.5mu}}

\begin{document}

\pagestyle{empty}
\hfill AEI-2014-013
\vskip 0.1\textheight
\begin{center}
{\Large{\bfseries 
	On conformal higher spin wave operators
}}
\vskip 0.1\textheight

{\bfseries Teake Nutma and Massimo Taronna} \\%

\vskip 0.5cm
\emph{\href{http://www.aei.mpg.de/}{Max-Planck-Institut f\"ur
Gravitationsphysik}}\\
\emph{(Albert Einstein Institut)}\\ 
\emph{Am M\"uhlenberg 1, } \\ 
\emph{14476 Golm, Germany}	\\ 

\vskip 0.5cm
{\tt \{teake.nutma, massimo.taronna\}@aei.mpg.de}%

\end{center}

\vskip 0.05\textheight

\begin{center} 
	{\bf Abstract }  
\end{center} 
\begin{quotation}
	\noindent  We analyze free conformal higher spin actions and the 
	corresponding wave operators in arbitrary even dimensions and backgrounds. 
	We show that the wave operators do not factorize in general, and identify 
	the Weyl tensor and its derivatives as the obstruction to factorization.
	We give a manifestly factorized form for them on (A)dS 
	backgrounds for arbitrary spin and on Einstein backgrounds for spin 2. 
	We are also able to fix the conformal wave operator in $d=4$ for $s=3$ up 
	to linear order in the Riemann tensor on generic Bach-flat backgrounds.
\end{quotation}

\newpage
\pagestyle{plain}

\tableofcontents

\section{Introduction}

Conformal gauge theories have received quite some attention over the years.  In
particular, the actions of Weyl gravity and conformal supergravity, together
with their corresponding wave equations, have been studied in great detail
\cite{Kaku:1978nz, Bergshoeff:1980is, Fradkin:1981jc, Fradkin:1982xc,
Deser:1983mm, Deser:1983tm, Riegert:1984hf, Fradkin:1985am, Boulanger:2001he,
Boulanger:2002bt,Boulanger:2004eh} as natural extensions of ordinary gravity and
supergravity theories. Interest has been also devoted to the corresponding
higher spin generalizations \cite{Fradkin:1990ps, Segal:2002gd,
Shaynkman:2004vu, Metsaev:2007rw, Vasiliev:2007yc, Vasiliev:2009ck, Florakis:2014kfa}, not just
because of the intriguing role of conformal symmetry. Flat space higher spin
(HS) fields are namely naturally endowed with higher derivative linearized
curvatures \cite{deWit:1979pe} that play a key role in conformal gauge
theories.\footnote{See \cite{Vasiliev:1999ba, Bekaert:2005vh, Bekaert:2010hw,
Sagnotti:2011qp, Didenko:2014dwa} for some reviews of HS theories.}

More recently, conformal HS fields have found interesting applications in the
context of the AdS/CFT correspondence. There, they play the role of sources to
the conformal currents, defined in the free O(N) vector models as well as in
generic CFT's in their free limit \cite{Leigh:2003ez, Metsaev:2008fs,
Metsaev:2009ym, Bekaert:2010ky, Joung:2011xb, Gover:2011rz, Vasiliev:2012vf, Giombi:2013yva}.

Nonetheless, it is important to keep in mind that HS conformal theories are
naturally higher derivative theories and for this reason violate unitarity, just
as conformal gravity. This feature allows them to bypass the Coleman-Mandula
theorem as well as other powerful no-go theorems in flat space.\footnote{See
e.g. \cite{Bekaert:2010hw} and references therein for a review of various no-go
theorems and \cite{Joung:2013nma} for a stronger version of the Coleman-Mandula
theorem in flat space.} On the other hand it has been recently pointed out how
asymptotically AdS solutions of Einstein gravity can be recovered from four
derivative theories by choosing appropriate boundary conditions
\cite{Metsaev:2007fq, Maldacena:2011mk, Lu:2011ks}. This provides some key hints
about the role of the latter non-unitary theories in the context of AdS/CFT.
Therefore, these features motivate a closer look at conformal HS theories and
their properties.

Free Lagrangians and the corresponding wave equations involving massless
Fronsdal fields and their variants have received considerable interest
\cite{Fronsdal:1978rb, Lopatin:1987hz, Pashnev:1998ti, Buchbinder:2001bs, Sagnotti:2003qa, Francia:2005bu, Skvortsov:2006at,
Buchbinder:2007ak, Skvortsov:2007kz, Francia:2008hd, Campoleoni:2012th}. But the
explicit form of the conformal wave operator for HS fields in curved spaces has
not been worked out yet.\footnote{See \cite{Joung:2012qy} for some discussion of
higher derivative theories in flat space, \cite{Erdmenger:1997gy,
Erdmenger:1997wy} for some earlier discussion on conformal operators and
\cite{branson1985, Graham, GJMS, Gover} for selected math literature.} The aim
of this paper  is to study free conformal higher spins actions and the
corresponding wave operators on generic backgrounds. One of our goals is to
discuss the factorization property of the conformal wave operator for HS fields
generalizing previous result for spin 2. We have also been able to fix the
conformal wave operator in $d=4$ for $s=3$ up  to linear order in the Riemann
tensor on generic Bach-flat backgrounds. As a byproduct of our analysis, we
obtain the full conformal wave operator on (A)dS backgrounds in any dimension in
a manifestly factorized form. Each factor turns out to be given by a two
derivative operator. Their combined mass spectrum comprises the massless and
partially-massless points plus massive points in higher dimensions
\cite{Deser:1983mm, Higuchi:1986py, Higuchi:1986wu, Deser:2001us, Deser:2001xr,
Skvortsov:2006at, Alkalaev:2011zv, Deser:2012qg, Bekaert:2013zya}. This provides
additional evidence for previous conjectures made in \cite{Joung:2012qy,
Tseytlin:2013jya} and extends them. In addition, we also identify the Weyl
tensor and its derivatives as the obstruction to factorization for spin $s>2$ on
generic backgrounds. Furthermore, we rediscover the well known factorization of
the conformal wave operator for spin 2 on Einstein backgrounds
\cite{Fradkin:1981jc, Fradkin:1982xc, Deser:1983tm, Deser:1983mm}, and  extend
it to arbitrary dimensions.

The obstruction to factorization for spin $s > 2$ can be interpreted as a
conformal reincarnation of the Aragone-Deser obstruction \cite{Aragone:1979hx}
for two derivative HS wave operators. Indeed, the crucial difference between
spin $2$ and HS fields is the explicit appearance of the Weyl tensor within the
gauge variation of the generic two derivative operators. On the contrary, any
contribution proportional to the Weyl tensor can be eliminated for spin $1$ and
$2$, making their wave operator factorizable.

Amongst other things we also develop a variant of the HS tractor calculus (see
e.g \cite{Joung:2013doa} and references therein), that finds potentially useful
applications to conformal HS fields. We believe that this formalism might
provide a useful tool for addressing various problems with conformal higher spin
fields, like for instance the extension of the present analysis to interactions
and to the study of conformal HS algebras.

The organization of the paper is as follows. In \autoref{Sec:Conformal_fields}
we describe a convenient formalism that allows us to deal with conformal fields
in a simplified way. In \autoref{Sec:Factorization Ansatz} we test the
consistency of a factorized conformal wave operator in generic backgrounds
studying the obstructions to factorization. In \autoref{sec:spin3} we give
the spin 3 conformal wave operator on Bach-flat backgrounds up to linear order
in the Riemann tensor. In \autoref{Sec:Conclusions} we
summarize our results and conclude. We have put additional
material that includes a discussion about gauge fixing and some lower-spin
examples in the appendices. Lastly, we have attached a Mathematica notebook 
containing independent checks as an ancillary file.

\section{Conformal higher spin fields}
\label{Sec:Conformal_fields}

Conformal higher spin fields \cite{Fradkin:1985am,Segal:2002gd} can be defined 
at the linear level by demanding the following gauge invariance properties
\begin{subequations}
\label{eq:gauge_trans}
\begin{align}
	\delta_\xi \, \varphi_{\mu_1 \cdots \mu_s} 
		& = \nabla_{(\mu_1} \xi_{\mu_2 \cdots \mu_s)}  , \label{eq:gauge_trans_fronsdal}\\
	\delta_\alpha \, \varphi_{\mu_1 \cdots \mu_s}  
		& = g_{(\mu_1 \mu_2} \alpha_{\mu_3 \cdots \mu_s)} .	\label{eq:gauge_trans_scale}
\end{align}
\end{subequations}
No trace constraints on fields or gauge parameters are imposed. The above
generalizes the linearized gauge invariance and rescaling invariance of
conformal gravity. Indeed, for spin 2 equation \eqref{eq:gauge_trans_scale}
describes linear dilatations  (scale transformations). For higher spins, on top
of the above transformations, one would in principle also need to consider also
proper HS scale transformations of the form $\varphi_{\mu_1 \cdots \mu_s}
\rightarrow \Omega \, \varphi_{\mu_1 \cdots \mu_s}$. But for the purpose of
this paper it will not be necessary to impose this beforehand. Irrespectively,
the wave operators we find turn out to be automatically invariant under these 
scale transformations.

We will now switch to an operator notation where fields are represented by 
generating functions,
\begin{equation}
\label{eq:genfunc}
	\varphi_{\mu_1 \cdots \mu_s}(x) \; \rightarrow \; \varphi(x,u) = 
		\frac{1}{s!}\varphi_{\mu_1 \cdots \mu_s}(x) e_{a_1}^{\ \mu_1}(x)
		u^{a_1} \cdots e_{a_s}^{\ \mu_s}(x)u^{a_s} .
\end{equation}
Here we have introduced a constant auxiliary tangent variable $u^a$. See 
\autoref{Appendix:Relations} for all our notational conventions and a brief 
introduction to the operator formalism. In the operator notation the gauge 
invariance properties \eqref{eq:gauge_trans} take the form
\begin{subequations}
\label{eq:gauge_trans_u}
\begin{align}
	\delta_\xi \varphi(x,u)
		& = u\cdot\nabla \xi(x,u) , \label{eq:gauge_trans_fronsdal_u}\\
	\delta_\alpha \varphi(x,u)
		& = u^2 \alpha(x,u) .	\label{eq:gauge_trans_scale_u}
\end{align}
\end{subequations}
From this it follows that a conformal field can be regarded as an equivalence 
class of standard massless higher spin fields defined on the cone $u^2\sim0$.
This observation allows us to use so-called Thomas-D derivatives 
$\hat\partial_u$ in the auxiliary variable $u$. Again, see 
\autoref{Appendix:Relations} for more information.

We now summarize our results. We find the following manifestly factorized form of
the spin $s$ conformal wave operator in $\text{(A)dS}_d$:
\begin{equation}
\label{eq:Conf_Operator}
	{\mathcal O}^{(s)} =
		\prod_{i=1}^{\tfrac{d}{2}-2+s} \left[
			\dal 
			- \tfrac{d-4+2s}{i(d-3-i+2s)}u\cdot\nabla\nabla\cdot\hat\partial_u
			+ \Lambda[(i-s+1)(i-s-d+2)-s]
		\right],
\end{equation}
Similarly, the factorized spin 2 conformal wave operator on any Einstein 
background can be expressed as
\begin{equation}
\label{eq:Spin2_Operator}
{\mathcal O}^{(2)} = 
	\prod_{i=1}^{\tfrac{d}{2}}\left[
		\dal
		- \tfrac{d}{i(d+1-i)}u\cdot\nabla\nabla\cdot\hat\partial_u 
		+ \Lambda[(i-1)(i-d)-2]
		+ W_{\mu\nu\rho\sigma}u^\mu u^\rho\hat\partial_{u_\nu}\hat\partial_{u_\rho}\right] .
\end{equation}
The conformal wave operator for higher spins does not factorize on generic 
Einstein spaces, as we shall demonstrate in the next section.

\section{Factorization of conformal wave operators}
\label{Sec:Factorization Ansatz}

In this section we study the obstructions for a factorized conformal wave
operator to be gauge invariant on generic backgrounds. Our soon to be disproved
assumption is that the conformal wave operator factorizes into two-derivative 
operators on any Bach-flat background, or generalizations thereof in $d>4$. 
The existence of a conformal wave operator on Bach-flat backgrounds can
be argued on the basis of the following two observations.

Firstly, an Aragone and Deser type of obstruction \cite{Aragone:1979hx} cannot
arise since the conformal coupling with gravity has the same number of
derivatives as the kinetic term at any order in the spin $s$ field. In
particular any coupling of the type $s-s-2-\ldots-2$ involving $n$ spin two
fields and two spin $s$ fields must involve $2s+d-4$ derivatives. This type of
obstruction appears for two derivative operators like the Fronsdal operator 
because the corresponding gravitational couplings are
higher derivative \cite{Fradkin:1987ks, Boulanger:2008tg, Joung:2013nma}.

Secondly, any possible tadpoles (i.e.~vertices linear in the higher spin field)  can
be removed. In principle one might be forced to add them, but they can always be
integrated by parts into a non-linear equation for the metric. In $d=4$ this
equation will involve the Bach tensor, although in general it will become a
higher derivative condition for the metric containing $2s+d-4$ derivatives. For
this  reason it will be compatible with, if not equivalent to, the conformal
gravity equations of motion.

We will come back to constructing a conformal invariant operators on
generic backgrounds in \autoref{sec:spin3}. But first we will concentrate on an
Ansatz that is explicitly factorized, with the aim of identifying the
obstruction to its gauge invariance. The non-existence of a factorized solution
in general will not imply the non-existence of the full operator. In fact, we
expect to the full operator to exist for any spin on generic conformal manifolds
for the reason mentioned above.

\subsection{Ansatz}

A ($2s+d-4$)-derivative factorized Ansatz for the conformal spin $s$ wave operator
can be written as
\begin{equation}
\label{eq:factorization_ansatz}
	{\mathcal O}^{(s)} = 
		\prod_{i=1}^{\tfrac{d}{2}-2+s}{\mathcal F}_i,
\end{equation}
where ${\mathcal F}_i$ is the most general Ansatz for a two derivative operator:
\begin{equation}
\label{eq:factor_ansatz}
{\mathcal F}_i = 
	\dal 
	+ \alpha_i u\cdot\nabla\nabla\cdot\hat\partial_u 
	+ \beta_i \Lambda 
	+ \gamma_i R^\Lambda_{\mu\nu\rho\sigma}u^\mu u^\rho\hat\partial_{u_\nu}\hat\partial_{u_\sigma} 
	+ \delta_i R^\Lambda_{\mu\nu}u^\mu\hat\partial_{u_\nu} 
	+\sigma_i R^\Lambda.
\end{equation}
Here we have defined $R^{\Lambda}_{\mu\nu\rho\sigma} = R_{\mu\nu\rho\sigma} -
\Lambda (g_{\mu\rho}g_{\nu\sigma} - g_{\nu\rho}g_{\mu\sigma})$, and similarly
for the Ricci tensor (see also \autoref{Appendix:Relations}). On Einstein
backgrounds this simplifies to
\begin{equation}
	{\mathcal F}^{\text E}_i = 
		\dal 
		+ \alpha_i u\cdot\nabla\nabla\cdot\hat\partial_u 
		+ \beta_i \Lambda 
		+ \gamma_i W_{\mu\nu\rho\sigma}u^\mu u^\rho\hat\partial_{u_\nu}\hat\partial_{u_\sigma},
\end{equation}
where $W_{\mu\nu\rho\sigma}$ is the Weyl tensor.

For the purpose of enforcing gauge invariance of the full operator it is useful
to compute the gauge transformation of the generic two-derivative operator
${\mathcal F}_i$. It reads
\begin{align}
\label{eq:Fgaugevar}
{\mathcal F} u\cdot\nabla = & \; 
	u\cdot\nabla\Bigl[
		(1+\alpha)\dal+\alpha\left(1-\tfrac{2}{h}\right)u\cdot\nabla\nabla\cdot\hat\partial_{u} \nonumber \\ 
&		\hspace{41.5pt}+\Lambda\left\{\beta\,+\,u\cdot\hat\partial_u+1\,+(\alpha u\cdot\hat\partial_u+1)(u\cdot\hat\partial_{u}+d-2)\right\}
	\Bigr]\nonumber\\
& 	- \alpha\left(u\cdot\nabla R^\Lambda_{\mu\nu\rho\sigma}\right)
		u^\mu u^\rho\hat\partial_{u_\nu}\hat\partial_{u_\sigma}
	+ (\gamma-\alpha)R^\Lambda_{\mu\nu\rho\sigma}
		u\cdot\nabla u^\mu u^\rho\hat\partial_{u_\nu}\hat\partial_{u_\sigma} \nonumber\\
&	+ 2(\gamma-1)R^\Lambda_{\mu\nu\rho\sigma}u^\mu\nabla^\nu u^\rho\hat\partial_{u_\sigma} \mspace{10.0mu}
	+ (1+\alpha)\left(u\cdot\nabla R^\Lambda_{\mu\nu}\right)u^\mu\hat\partial_{u_\nu} \nonumber\\
&	+ \; (\alpha+\delta)\;R^\Lambda_{\mu\nu}u\cdot\nabla u^\mu\hat\partial_{u_\nu} \mspace{25mu}
	+ (1+\delta)R^\Lambda_{\mu\nu}u^\mu\nabla^\nu
	- u^\mu u^\nu\left(\nabla\cdot\hat\partial_u R^\Lambda_{\mu\nu}\right) \nonumber\\
&	- \tfrac{2}{h-2}(\gamma+\delta)R^\Lambda_{\mu\nu}u^\mu u^\nu\nabla\cdot\hat\partial_u \mspace{7.0mu}
	+ \sigma R^\Lambda u\cdot\nabla .
\end{align}
First of all, the structure of the gauge variation illustrates an important
difference between spin-$2$ and higher spins. This is due to the appearance of
terms proportional to the full Riemann tensor and its derivatives, for
instance
\begin{equation}
	\left(u\cdot\nabla R^\Lambda_{\mu\nu\rho\sigma}\right) 
		u^\mu u^\rho\hat\partial_{u_\nu}\hat\partial_{u_\sigma}.
\end{equation}
This term, being cubic in the auxiliary variable $u$, appears only for spin 
$s\geq3$. This is actually a reincarnation of the same feature pointed out by 
Aragone and Deser \cite{Aragone:1979hx} in the context of Fronsdal fields. 

The above gauge variation can be used to recursively compute the gauge variation 
of the factorized Ansatz \eqref{eq:factorization_ansatz}. Using the notation
\begin{equation}
	{\mathcal F}_i u\cdot\nabla = 
		u\cdot\nabla \tilde {\mathcal F}_i + {\mathcal X}_i,
\end{equation}
we get
\begin{equation}
	\left({\mathcal F}_1\cdots {\mathcal F}_n \right) u\cdot\nabla = 
		u\cdot\nabla\left(\tilde{\mathcal F}_1\cdots \tilde{\mathcal F}_n \right)+{\mathcal X}^{(n)},
\end{equation}
where ${\mathcal X}^{(n)}$ is recursively defined as
\begin{equation}
	{\mathcal X}^{(n)} 
		= {\mathcal X}_1 \tilde{\mathcal F}_{2}\cdots \tilde{\mathcal F}_n 
			+ {\mathcal F}_1 {\mathcal X}^{(n-1)}
		= \sum_{k=1}^n 
			{\mathcal F}_{1}\cdots {\mathcal F}_{k-1} \, {\mathcal X}_k \,
			\tilde{\mathcal F}_{k+1}\cdots \tilde{\mathcal F}_n.
\end{equation}
It is then straightforward to see that gauge invariance for the spin $s$ wave 
operator implies the condition
\begin{align}
\label{eq:gaugeinvarcond}
	u\cdot\nabla\left(\tilde{\mathcal F}_{\tfrac{d}{2}-2+s}\cdots \tilde{\mathcal F}_1 \right)
	+ {\mathcal X}^{\left(\tfrac{d}{2}-2+s\right)} 
		= 0.
\end{align}

\subsection{Arbitrary spins on AdS backgrounds}\label{Sec:Zeroth_Order}

The coefficients $\alpha$ and $\beta$ enter the Riemann-independent part of the
Ansatz \eqref{eq:factor_ansatz}. Hence, in order to fix them it is
sufficient to look at the zeroth order in the $R^\Lambda$ tensors. This
corresponds to the case of (A)dS, which we are going to consider in detail in
this section. The equation \eqref{eq:gaugeinvarcond} will simplify and will admit an
iterative structure, which is crucial for  factorization.

At zeroth order in Riemann tensors we have ${\mathcal X}_i\sim0$, and end up with
the following equation for the $i^\textrm{th}$ factor in the Ansatz:
\begin{multline}
	\left(\nabla\cdot\hat\partial_u\right)^{i-1}
	\biggl[
		(1+\alpha_i)\dal+\alpha_i(1-\tfrac{2}{h})u\cdot\nabla\nabla\cdot\hat\partial_{u}\\
		+\Lambda\left(
			\beta_i+u\cdot\hat\partial_u+
			1+(\alpha_i u\cdot\hat\partial_u+1)(u\cdot\hat\partial_{u}+d-2)
		\right)
	\biggr]
	\sim \left(\nabla\cdot\hat\partial_u\right)^i .
\end{multline}
This recursively ensures that all terms proportional to
$\left(\nabla\cdot\hat\partial_u\right)^{i-1}$ vanish. In principle we should
also impose that no higher divergence is generated, but this condition turns out
to be automatically satisfied if the number of derivatives is chosen to be
$2s+d-4$. We will now fix all $\alpha$'s and $\beta$'s by solving linear
equations. We begin with observing that
\begin{equation}
\label{eq:multidivcomm}
	\left(\nabla\cdot\hat\partial_u\right)^k \tilde{\mathcal F}_i =
		\left(\nabla\cdot\hat\partial_u\right)^{k-i} \left[
			a_i\dal 
			+ b_i u\cdot\nabla \nabla\cdot\hat\partial_u
			+ c_i\Lambda
		\right] (\nabla\cdot\hat\partial_u)^{i} 
		+ \mathcal{O}\left({R^\Lambda} \right),
\end{equation}
where the coefficients satisfy the following recursion relations:
\begin{subequations}
\begin{align}
	a_j &= a_{j-1}+b_{j-1},\\
	b_j &= b_{j-1}\left(1-\frac{1}{\tfrac{d}{2}-2+s-j}\right),\label{eq:bs}\\
	c_j &= c_{j-1}+b_{j-1}(s-j-1)(s-j+d-3)+a_{j-1}\bigl(2(s-j)+d-3\bigr).
\end{align}
\end{subequations}
These recursion relations have boundary conditions
\begin{subequations}
\begin{align}
	a_0 &= 1+\alpha_i,\\
	b_0 &= \alpha_i\left(1-\frac{1}{\tfrac{d}{2}-2+s}\right),\\
	c_0 &= \beta_i+s+(\alpha_i(s-1)+1)(s+d-3).
\end{align}
\end{subequations}
The solution to the first two recursion relations reads:
\begin{subequations}
\begin{align}
	a_j &= 1+\alpha_i\left[1+j\left(1-\frac{1+j}{d-4+2s}\right)\right],\\
	b_j &= \alpha_i\left(1-\frac{1+j}{\tfrac{d}{2}-2+s}\right).
\end{align}
\end{subequations}
We do not write the solution for $c_i$ since it is rather cumbersome and enters
only intermediate steps of the computation.
We can now enforce gauge invariance by recursively demanding that terms 
proportional to a divergence vanish in the gauge variation.
We end up with
\begin{subequations}
\begin{align}
	a_{i-1}(\alpha_i)			& = 0,\\
	c_{i-1}(\alpha_i,\beta_i)	& = 0.
\end{align}
\end{subequations}
The solution to these equations reads
\begin{subequations}
\label{eq:alphabetasolution}
\begin{align}
	\alpha_i &= -\frac{1}{1+(i-1)\left(1-\tfrac{i}{d-4+2s}\right)},\\
	\beta_i  &= (i-s+1)(i-s-d+2)-s.
\end{align}
\end{subequations}
And so the conformal wave operator on (A)dS takes the form
\begin{equation}
{\mathcal O}^{(s)} = 
	\prod_{i=1}^{\tfrac{d}{2}-2+s}\left[
		\dal
		-\frac{d-4+2s}{i(d-3-i+2s)}u\cdot\nabla\nabla\cdot\hat\partial_u
		+\Lambda\bigl((i-s+1)(i-s-d+2)-s\bigr)
	\right].
\end{equation}
Finally, the action which has ${\mathcal O}^{(s)}\varphi_{(s)} = 0$ as an equation 
of motion reads
\begin{equation}
	S^{(s)} = \tfrac{1}{2} \int d^dx \sqrt{-g}\ e^{\hat\partial_{u_1}\cdot\hat\partial_{u_2}}\varphi_{(s)}(u_1) {\mathcal O}^{(s)}\varphi_{(s)}(u_2)\bigg|_{u_i=0}.
\end{equation}
This reproduces the correct equations of motion because the operator 
${\mathcal O}$ is automatically self-adjoint up to total derivatives.

It is worth pointing out that the coefficients $\beta$ precisely match the
masses associated with the partially massless points for spin $s$, plus some
discrete massive points in $d>4$. This is in agreement with previous conjectures
on conformal HS wave operators \cite{Joung:2012qy,Tseytlin:2013jya}. This
implies in turn that the part of the conformal operator that is \emph{not}
proportional to divergences or traces has the form
\begin{equation}\label{eq:eomTT}
	{\mathcal O} \sim
		\prod_{i=0}^{\tfrac{d}{2}-3+s}\Bigl[
			\dal+\Lambda \bigl((i-s+2)(i-s-d-3)-s\bigr)
		\Bigr].
\end{equation}
In \autoref{Appendix:ConformalWaveNonF} we show that terms involving divergences
can be set to zero by choosing a convenient gauge.

Before concluding this section it is important to comment that strictly speaking
the above discussion is sufficient to determine the full conformal spin $s$
operator on (A)dS only in $d=4$, where the number of derivatives required by
scale invariance is $2s$.  In higher even dimensions the first $s$ factors have
to be the same as above but the next $\tfrac{d}{2}-2$ factors are not
constrained by gauge invariance and one would need to analyze conformal
invariance more closely. Notice that conformal invariance,
\begin{equation}
\label{eq:conformaltransfo}
	\delta g_{\mu\nu}=\Omega(x)^2g_{\mu\nu},
\end{equation}
is not easy to prove
due to the generically complicated transformation properties of covariant derivatives. However, the condition of gauge
invariance at the operator level,
\begin{equation}
\label{eq:Ogauge}
	{\mathcal O}^{(s)}\,u\cdot\nabla = 0,
\end{equation}
is strong enough to completely fix the wave
operator in any dimension. Loosely speaking, operator gauge invariance means that
the kernel of ${\mathcal O}^{(s)}\,u\cdot\nabla$ is enlarged from the HS gauge parameters $\xi$ to 
arbitrary homogeneous functions of $u$.\footnote{Enlarging the domain of formal generating
functions to distributions has also been done in \cite{Gelfond:2013xt}.} 
Moreover, we have also checked in \autoref{Appendix:ConformalWaveNonF} 
that the factorization Ansatz does not play any role and one can arrive
at analogous results starting from a more general Ansatz. One can then
argue, and check with examples (see \autoref{Appendix:Examples}), that the
stronger gauge invariance condition \eqref{eq:Ogauge},
implies conformal invariance when the operator ${\mathcal O}_s$ is defined on
the equivalence classes \eqref{eq:confeq}. In $d=4$ the crucial simplification
is that the operator gauge invariance and the usual gauge invariance conditions
coincide. 

From a group-theoretical perspective the operator gauge invariance implies also that
the pattern of masses follows a very simple relation, 
\begin{equation}
	E=d+s-3-i \quad \textrm{for } 0\leq i\leq \tfrac{d}{2}-3+s .
\end{equation} 
This is nothing but the continuation of the pattern of the (partially-)massless
points, $0\leq i\leq s-1$, to massive points. Furthermore, it is what is expected from the
decomposition of a representation of the conformal algebra with respect to the
(A)dS subalgebra \cite{Tseytlin:2013jya}. It is remarkable that the above
requirements can be recast in terms of a usual gauge invariance condition
extended to the operator level. For these reasons, it might provide a useful tool to control conformal
invariance \eqref{eq:conformaltransfo}.

So far we have been able to completely fix the conformal wave operator on (A)dS. In
the following we will analyze the same problem in generic backgrounds. We shall
first consider the spin-$2$ case in more detail, and then address the higher spin
problem.

\subsection{Spin 2 on generic backgrounds}

The spin $2$ case is special with respect to its higher spin cousins because the
commutation relations \eqref{eq:operatoralgebra2} simplify. In particular, terms
of order $u^3$ or $\hat\partial_u^3$ in the conformal operator as well as terms
of order $\hat\partial_u^2$ in gauge variation drop out. The gauge variation
of a single $\mathcal{F}$ \eqref{eq:Fgaugevar} simplifies to
\begin{subequations}
\begin{align}
	{\mathcal F} u\cdot\nabla = & \;
		u\cdot\nabla\Bigl[
			(1+\alpha)\dal+\alpha(1-\tfrac{2}{h})u\cdot\nabla\nabla\cdot\hat\partial_{u} \nonumber\\
& 			\hspace{40pt}+\Lambda\left(
				\beta
				+u\cdot\hat\partial_u+1
				+(\alpha u\cdot\hat\partial_u+1)(u\cdot\hat\partial_{u}+d-2)
			\right)
		\Bigr] \nonumber\\
&		+2(\gamma-1)R^\Lambda_{\mu\nu\rho\sigma}u^\mu\nabla^\nu u^\rho\hat\partial_{u_\sigma}
		+(1+\alpha)\left(u\cdot\nabla R^\Lambda_{\mu\nu}\right)u^\mu\hat\partial_{u_\nu} \nonumber\\
&		+\;(\alpha+\delta)\;R^\Lambda_{\mu\nu}u\cdot\nabla u^\mu\hat\partial_{u_\nu} \mspace{15mu}
		-u^\mu u^\nu\left(\nabla\cdot\hat\partial_u R^\Lambda_{\mu\nu}\right)
		+(1+\delta)R^\Lambda_{\mu\nu}u^\mu\nabla^\nu \nonumber\\
&		-\tfrac{2}{h-2}(\gamma+\delta)R^\Lambda_{\mu\nu}u^\mu u^\nu\nabla\cdot\hat\partial_u
		+\sigma R^\Lambda u\cdot\nabla .
\end{align}
\end{subequations}
This enables us to eliminate all instances of the Riemann tensor by simply 
choosing $\gamma=1$. This very simple observation is sufficient to ensure that
the factorized Ansatz works on any Einstein background. 

However, it should be
clear from the argument itself that this simplification is non-generic. For
completeness and to underline the non-generic nature, let us analyze the factorization of the conformal spin 2
operator on general backgrounds more closely. Taking the solution
\eqref{eq:alphabetasolution} for $\alpha$'s and $\beta$'s obtained in the
previous section into account, the gauge-invariance condition reads in $d=4$
\begin{equation}
	{\mathcal X}_1\tilde{\mathcal F}_2+{\mathcal F}_1{\mathcal X}_2=0 .
\end{equation}
Here we have
\begin{align}
	{\mathcal X}_i = &
		+(1+\alpha_i)\left(u\cdot\nabla R^\Lambda_{\mu\nu}\right)u^\mu\hat\partial_{u_\nu}
		\;\;+2(\gamma_i-1)R^\Lambda_{\mu\nu\rho\sigma}u^\mu\nabla^\nu u^\rho\hat\partial_{u_\sigma} \nonumber\\
&		+(\alpha_i+\delta_i)R^\Lambda_{\mu\nu}u\cdot\nabla u^\mu\hat\partial_{u_\nu}
		\quad\;\;+(1+\delta_i)R^\Lambda_{\mu\nu}u^\mu\nabla^\nu
		-u^\mu u^\nu\left(\nabla\cdot\hat\partial_u R^\Lambda_{\mu\nu}\right)
		 \nonumber\\
&		-\tfrac{2}{h-2}(\gamma_i+\delta_i)R^\Lambda_{\mu\nu}u^\mu u^\nu\nabla\cdot\hat\partial_u
		+\sigma_i R^\Lambda u\cdot\nabla\\
	{\mathcal F}_1 = & 
		\phantom{\tfrac13}\!\dal-u\cdot\nabla\nabla\cdot\hat\partial_u-2\Lambda\\
	\tilde{\mathcal F}_2 = &
		\tfrac13 \!\dal-\tfrac13 u\cdot\nabla\nabla\cdot\hat\partial_u-\Lambda.
\end{align}
The terms linear in $R_{\mu\nu\rho\sigma}^\Lambda$ without any divergence are
\begin{equation}
	2R_{\mu\nu\rho\sigma}^\Lambda u^\mu\nabla^\nu u^\rho\hat\partial_{u_\sigma}\Big[
		(\gamma_1-1)(\tfrac{1}{3} \dal-\Lambda)+(\gamma_2-1)(\dal-2\Lambda)
	\Big] .
\end{equation}
It is easy to see that the only solution to gauge invariance is $\gamma_i=1$,
which eliminates any instance of the Riemann tensor in the gauge variation. In
order to study the obstructions related to $R^\Lambda_{\mu\nu}$ it is useful to
first concentrate on the terms that do not involve any derivative of
$R^\Lambda_{\mu\nu}$. Thus for the moment we will set $\nabla_{\alpha}
R_{\mu\nu}\sim0$ and, as a consequence of the Bianchi identity, $R^\Lambda\sim0$
(i.e.~the non-constant part of the Ricci scalar vanishes). Dropping terms
proportional to divergences for simplicity, we get the following gauge
variation:
\begin{align}
	{\mathcal O}^{(2)} u \cdot \nabla \sim &
	+ \Big[(\delta_1+1)R^\Lambda_{\mu\nu}u^\mu\nabla^\nu+(\delta_1-1) R^\Lambda_{\mu\nu}u\cdot\nabla u^\mu\hat\partial_{u_\nu}\Big][\tfrac13 \dal-\Lambda] \nonumber\\
	& -\tfrac16 (\delta_1+1)R^\Lambda_{\mu\nu}u^\mu u^\nu\nabla\cdot\hat\partial_u \dal \nonumber\\
	& + [\dal-2\Lambda]\Big[(1+\delta_2)R^\Lambda_{\mu\nu}u^\mu\nabla^\nu+(\alpha_2+\delta_2)R^\Lambda_{\mu\nu}u\cdot\nabla u^\mu\hat\partial_{u_\nu}\Big]\nonumber\\
	& -u\cdot\nabla\nabla\cdot\hat\partial_u\Big[(1+\delta_2)R^\Lambda_{\mu\nu}u^\mu\nabla^\nu+(\alpha_2+\delta_2)R^\Lambda_{\mu\nu}u\cdot\nabla u^\mu\hat\partial_{u_\nu}\Big] .
\end{align}
Keeping only terms of the order $(R^\Lambda)^2$ and commuting all boxes until
they act on the gauge parameter while dropping divergences, we obtain:
\begin{align}
-\tfrac16 (\delta_1+1)R^\Lambda_{\mu\nu}u^\mu u^\nu R^\Lambda_{\mu\nu}\nabla^\mu\hat\partial_{u_\nu}&\nonumber\\
(1+\delta_2)\Big[-2R^\Lambda_{\mu\alpha} R^{\Lambda\,\alpha}_{\quad\,\nu\rho\sigma}u^\mu\nabla^\nu u^\rho\hat\partial_{u_\sigma}+R^\Lambda_{\mu\alpha}R^{\Lambda\,\alpha}_{\quad\,\nu}u^\mu\nabla^\nu\Big]&\nonumber\\
+(\alpha_2+\delta_2)\Big[-2R^\Lambda_{\mu\nu\rho\alpha}R^{\Lambda\,\alpha}_{\quad\,\sigma}u^\mu\nabla^\nu u^\rho 
\hat\partial_{u_\sigma}+R^\Lambda_{\mu\nu}u^\mu\nabla^\nu R^\Lambda_{\rho\sigma} u^\rho\hat\partial_{u_\sigma}\Big]&\nonumber\\
-(\alpha_2+\delta_2)u\cdot\nabla R^\Lambda_{\mu\beta}R^{\Lambda\,\beta}_{\quad\,\nu}u^\mu\hat\partial_{u_\nu}&.
\end{align}
This cannot be set to zero by tuning the free coefficients, which implies $R^\Lambda_{\mu\nu}$ is an obstruction to factorization in the spin two case.
This concludes the proof that factorization of the spin-$2$ conformal wave operator is possible only on Einstein backgrounds. As we have seen above its form is remarkably simple and can be written as
\begin{multline}
	{\mathcal O}^{(2)} = 
		\left(
			\dal 
			- u\cdot\nabla\nabla\cdot\hat\partial_u 
			-2 \Lambda 
			+ W_{\mu\nu\rho\sigma}u^\mu u^\rho\hat\partial_{u_\nu}\hat\partial_{u_\sigma}
		\right) \\
		\times \left(
			\dal 
			- \tfrac23 u\cdot\nabla\nabla\cdot\hat\partial_u 
			-4 \Lambda 
			+ W_{\mu\nu\rho\sigma}u^\mu u^\rho\hat\partial_{u_\nu}\hat\partial_{u_\sigma}
		\right).
\end{multline}
On more general conformal manifolds factorization is not possible. 

The above discussion generalizes readily to any dimension, upon which we get the 
following manifestly factorized form of the spin $2$ conformal wave operator:
\begin{equation}
{\mathcal O}^{(2)} = 
	\prod_{i=1}^{\tfrac{d}{2}}\left[
		\dal
		-\tfrac{d}{i(d+1-i)}u\cdot\nabla\nabla\cdot\hat\partial_u
		+\Lambda[(i-1)(i-d)-2]
		+W_{\mu\nu\rho\sigma}u^\mu u^\rho\hat\partial_{u_\nu}\hat\partial_{u_\sigma}
	\right],
\end{equation}

Before concluding this section, let us point out that the above result is  the
unique operator that factorizes, and it reduces to our previous result
\eqref{eq:Conf_Operator} upon restricting to (A)dS backgrounds. If the
factorization requirement is dropped more conformal operators can be found,
e.g.~by linearizing the conformal invariant densities of
\cite{Bastianelli:2000rs, Boulanger:2004zf, Metsaev:2010kp, Oliva:2010zd}.
However, all but one of these densities vanish when linearized on (A)dS
backgrounds as they consist of more than two Weyl tensors. See also
\autoref{sec:spin2data} for an example of this for $d=6$.

We will now proceed to the higher spin cases. Due to the generic nature of the
obstructions we found for spin $2$, we will restrict our attention to Einstein
manifolds in what follows.

\subsection{Higher spins on Einstein backgrounds}

We will now consider arbitrary spins on Einstein backgrounds, and consequently
set $R^\Lambda_{\mu\nu}$ to zero. Upon doing so, the commutation
relations simplify drastically and the gauge variation of a single
$\mathcal{F}$, equation \eqref{eq:Fgaugevar}, becomes
\begin{align}
\label{eq:FgaugeonEinstein}
	{\mathcal F} u\cdot\nabla & =
		u\cdot\nabla\left[
			(1+\alpha)\dal
			+\alpha(1-\tfrac{2}{h})u\cdot\nabla\nabla\cdot\hat\partial_{u}\right. \nonumber\\
			&\left.\hspace{40pt}
			+\Lambda\left(\beta+u\cdot\hat\partial_u+1+(\alpha u\cdot\hat\partial_u+1)(u\cdot\hat\partial_{u}+d-2)\right)
		\right] \nonumber\\
& 		-\alpha(u\cdot\nabla W_{\mu\nu\rho\sigma})u^\mu u^\rho\hat\partial_{u_\nu}\hat\partial_{u_\sigma}
		+(\gamma-\alpha)W_{\mu\nu\rho\sigma}u\cdot\nabla u^\mu u^\rho\hat\partial_{u_\nu}\hat\partial_{u_\sigma} \nonumber\\
&		+2(\gamma-1)W_{\mu\nu\rho\sigma}u^\mu\nabla^\nu u^\rho\hat\partial_{u_\sigma}.
\end{align}
To analyze if the Weyl tensor is an obstruction it is useful to drop all of its
derivatives and set
\begin{subequations}
\label{eq:DerWeylToZero}
\begin{align}
	\nabla_{\alpha}W_{\mu\nu\rho\sigma} &
		\sim 0,\\
	[\nabla_\beta, \nabla_\alpha] W_{\mu\nu\rho\sigma}&
		\sim 0 .
\end{align}
\end{subequations}
We can then rewrite equation \eqref{eq:FgaugeonEinstein} as
\begin{align}
{\mathcal F} u\cdot\nabla &\sim u\cdot\nabla\left[(1+\alpha)\dal+\alpha(1-\tfrac{2}{h})u\cdot\nabla\nabla\cdot\hat\partial_{u}\right.\\
				&\hspace{40pt}+\Lambda(\beta+u\cdot\hat\partial_u+1+(\alpha u\cdot\hat\partial_u+1)(u\cdot\hat\partial_{u}+d-2))
				\nonumber\\
				&\hspace{40pt}\left.+(\gamma-\alpha)W_{\mu\nu\rho\sigma} u^\mu u^\rho\hat\partial_{u_\nu}\hat\partial_{u_\sigma}\right]\nonumber\\
				&+2(\gamma-1)W_{\mu\nu\rho\sigma}u^\mu\nabla^\nu u^\rho\hat\partial_{u_\sigma}.\nonumber
\end{align}
The gauge variation of the factorized Ansatz becomes
\begin{equation}
	\delta {\mathcal O}^{(s)} = 
		u\cdot\nabla \tilde{\mathcal F}_1\cdots \tilde{\mathcal F}_{\tfrac{d}{2}-2+s}
		+\sum_{k=1}^{\tfrac{d}{2}-2+s} 
			{\mathcal F}_{1}\cdots {\mathcal F}_{k-1} 
			{\mathcal X}_k\tilde
			{\mathcal F}_{k+1}\cdots \tilde{\mathcal F}_{\tfrac{d}{2}-2+s},
\end{equation}
where
\begin{subequations}
\begin{align}
	\tilde{\mathcal F}_i & =
		(1+\alpha_i)\dal+\alpha_i(1-\tfrac{2}{h})u\cdot\nabla\nabla\cdot\hat\partial_{u} \nonumber \\
&\ \ 		+\Lambda\left(\beta_i+u\cdot\hat\partial_u
			+1+(\alpha_i u\cdot\hat\partial_u+1)(u\cdot\hat\partial_{u}+d-2)\right) \nonumber\\
&\ \ 		+(\gamma_i-\alpha_i)W_{\mu\nu\rho\sigma} u^\mu u^\rho\hat\partial_{u_\nu}\hat\partial_{u_\sigma}\\
	{\mathcal F}_i & = 
		\dal 
		+ \alpha_i u\cdot\nabla\nabla\cdot\hat\partial_u 
		+ \beta_i \Lambda 
		+ \gamma_i W_{\mu\nu\rho\sigma}u^\mu u^\rho\hat\partial_{u_\nu}\hat\partial_{u_\sigma}\\
	{\mathcal X}_k & = 
		2(\gamma_k-1)W_{\mu\nu\rho\sigma}u^\mu\nabla^\nu 
		u^\rho\hat\partial_{u_\sigma}.
\end{align}
\end{subequations}
We can now concentrate on terms involving the Weyl tensor via the combination
\begin{equation}
	\left(W_{\mu\nu\rho\sigma}u^\mu u^\rho\hat\partial_{u_\nu}\hat\partial_{u_\sigma}\right)^m.
\end{equation}
These include terms proportional to powers of the Weyl tensor and the gauge parameter $\xi$,
\begin{equation}
	W_{\mu_1\ \nu_1}^{\ \alpha_1\ \beta_1}
	W_{\alpha_1\ \beta_1}^{\ \alpha_2\ \beta_2}\cdots 
	W_{\alpha_m\ \beta_m}^{\ \rho_1\ \sigma_1}
	\,\xi_{\rho_1\sigma_1\ldots},
\end{equation}
and are non vanishing upon setting the derivatives of the Weyl tensor to zero. 
Moreover, they can arise only from the first contribution to the gauge variation.
For this reason they need to vanish identically, so we are forced to impose the
following condition:
\begin{equation}
	\gamma_i=\alpha_i,\quad\forall\ i.
\end{equation}
Notice that we have used the defining properties of the $\alpha$'s and $\beta$'s
in eq.~\eqref{eq:multidivcomm} to simplify the terms involving divergences.
However, when we now shift our attention to terms that involve the Weyl tensor
via the combination
\begin{equation}
	\left(
		W_{\mu\nu\rho\sigma}u^\mu u^\rho\hat\partial_{u_\nu}\hat\partial_{u_\sigma}
	\right)^{m-1} 
	W_{\mu\nu\rho\sigma}u^\mu\nabla^\nu u^\rho\hat\partial_{u_\sigma},
\end{equation}
we see that they do not vanish for covariantly constant Weyl tensors. 
Thus gauge invariance also requires
\begin{equation}
\gamma_i=1,\quad\forall\ i.
\end{equation}
The above clash of the gauge invariance condition identifies these particular
Weyl tensor combinations, and hence generically the Weyl tensor, as the generic
obstruction to factorization for the spin $s$ conformal wave operator on
Einstein backgrounds. Moreover, we can also identify the first derivative of the
Weyl tensor as an independent obstruction to factorization. This can be seen
from \eqref{eq:FgaugeonEinstein} by looking at the contributions proportional to
$\alpha(u\cdot\nabla W_{\mu\nu\rho\sigma}u^\mu
u^\rho\hat\partial_{u_\nu}\hat\partial_{u_\sigma})$, since none of the $\alpha$'s
is vanishing.

We have performed various independent checks of the above computations
explicitly with the help of Mathematica. We have attached the corresponding notebook
to this paper where the explicit spin 3 wave operator has been constructed up 
linear order in the Riemann tensor. In the next section we briefly summarize
the contents of the notebook.

\section{Spin 3 wave operator on Bach-flat backgrounds}
\label{sec:spin3}

With the help of Mathematica we have worked out the explicit form of the  unique
spin 3  conformal wave operator in $d=4$ up to linear terms in the Riemann
tensor on Bach-flat backgrounds. We have done this by simply listing all
possible contractions and constructing a gauge invariant Ansatz out of those. As
expected from our  arguments in \autoref{Sec:Factorization Ansatz}, we did not
find any obstruction.

Furthermore, we also confirmed the invariance of the wave operator under Weyl
rescalings of the metric \eqref{eq:conformaltransfo}. Remarkably, this turned 
out to be automatically the case after
imposing gauge invariance under \eqref{eq:gauge_trans}.

Even at linear order in Riemann tensors, the wave operator is rather unwieldy,
consisting of roughly 200 terms. Its full form can be found in the attached notebook.
Here we present the wave operator on Ricci flat backgrounds. It reads:
\begin{align}
\mathcal{O}^{(3)}_{\mu\nu\rho}(\varphi) = &
- \tfrac{21}{10} \nabla_{\mu \tau}R_{\nu \alpha \rho \beta} \nabla^{\sigma}{}_{\sigma}\hat\varphi^{\tau \alpha \beta} 
-  \tfrac{7}{10} \nabla_{\nu \alpha}R_{\rho \sigma \tau \beta} \nabla^{\sigma \tau}\hat\varphi_{\mu}{}^{\alpha \beta} 
+ \tfrac{182}{25} \nabla_{\sigma \tau}R_{\nu \alpha \rho \beta} \nabla^{\sigma \tau}\hat\varphi_{\mu}{}^{\alpha \beta} 
\nonumber\\ &
-  \tfrac{49}{25} \nabla_{\mu \nu}R_{\rho \alpha \tau \beta} \nabla^{\sigma \tau}\hat\varphi_{\sigma}{}^{\alpha \beta} 
-  \tfrac{49}{25} \nabla_{\mu \tau}R_{\nu \alpha \rho \beta} \nabla^{\sigma \tau}\hat\varphi_{\sigma}{}^{\alpha \beta} 
- 7 \nabla_{\mu \alpha}R_{\nu \tau \rho \beta} \nabla^{\sigma \tau}\hat\varphi_{\sigma}{}^{\alpha \beta} 
\nonumber\\ &
-  \tfrac{259}{25} \nabla_{\mu}{}^{\sigma}\hat\varphi^{\tau \alpha \beta} \nabla_{\tau \sigma}R_{\nu \alpha \rho \beta} 
-  \tfrac{84}{25} \nabla_{\mu}{}^{\sigma}\hat\varphi^{\tau \alpha \beta} \nabla_{\tau \alpha}R_{\nu \sigma \rho \beta} 
+ \tfrac{721}{50} \nabla^{\sigma \tau}\hat\varphi_{\mu}{}^{\alpha \beta} \nabla_{\alpha \tau}R_{\nu \sigma \rho \beta} 
\nonumber\\ &
-  \tfrac{161}{50} \nabla^{\sigma \tau}\hat\varphi_{\mu}{}^{\alpha \beta} \nabla_{\alpha \beta}R_{\nu \sigma \rho \tau} 
-  \tfrac{21}{5} \nabla_{\mu}{}^{\sigma}\hat\varphi_{\sigma}{}^{\tau \alpha} \nabla^{\beta}{}_{\beta}R_{\nu \tau \rho \alpha} 
+ \tfrac{252}{25} \nabla^{\sigma \tau}\hat\varphi_{\mu \sigma}{}^{\alpha} \nabla^{\beta}{}_{\beta}R_{\nu \tau \rho \alpha} 
\nonumber\\ &
-  \tfrac{35}{2} \nabla^{\sigma \tau}\hat\varphi_{\mu \nu}{}^{\alpha} \nabla^{\beta}{}_{\beta}R_{\rho \sigma \tau \alpha} 
+ \tfrac{343}{50} \nabla_{\mu}{}^{\sigma}\hat\varphi_{\nu}{}^{\tau \alpha} \nabla^{\beta}{}_{\beta}R_{\rho \tau \sigma \alpha} 
-  \tfrac{7}{5} \nabla_{\tau}R_{\rho \alpha \sigma \beta} \nabla_{\mu \nu}{}^{\sigma}\hat\varphi^{\tau \alpha \beta} 
\nonumber\\ &
+ \tfrac{7}{50} \nabla^{\sigma}\hat\varphi^{\tau \alpha \beta} \nabla_{\mu \nu \tau}R_{\rho \alpha \sigma \beta} 
-  \tfrac{42}{5} \nabla_{\tau}R_{\nu \alpha \rho \beta} \nabla_{\mu}{}^{\sigma}{}_{\sigma}\hat\varphi^{\tau \alpha \beta} 
+ \tfrac{161}{50} \nabla_{\rho}R_{\sigma \alpha \tau \beta} \nabla_{\mu}{}^{\sigma \tau}\hat\varphi_{\nu}{}^{\alpha \beta} 
\nonumber\\ &
+ \tfrac{399}{50} \nabla_{\tau}R_{\rho \alpha \sigma \beta} \nabla_{\mu}{}^{\sigma \tau}\hat\varphi_{\nu}{}^{\alpha \beta} 
+ \tfrac{441}{50} \nabla_{\alpha}R_{\rho \sigma \tau \beta} \nabla_{\mu}{}^{\sigma \tau}\hat\varphi_{\nu}{}^{\alpha \beta} 
- 7 \nabla_{\nu}R_{\rho \alpha \sigma \beta} \nabla_{\mu}{}^{\sigma \tau}\hat\varphi_{\tau}{}^{\alpha \beta} 
\nonumber\\ &
-  \tfrac{42}{5} \nabla_{\sigma}R_{\nu \alpha \rho \beta} \nabla_{\mu}{}^{\sigma \tau}\hat\varphi_{\tau}{}^{\alpha \beta} 
-  \tfrac{49}{5} \nabla_{\alpha}R_{\nu \sigma \rho \beta} \nabla_{\mu}{}^{\sigma \tau}\hat\varphi_{\tau}{}^{\alpha \beta} 
-  \tfrac{203}{50} \nabla^{\sigma}\hat\varphi^{\tau \alpha \beta} \nabla_{\mu \tau \sigma}R_{\nu \alpha \rho \beta} 
\nonumber\\ &
-  \tfrac{21}{10} \nabla^{\sigma}\hat\varphi^{\tau \alpha \beta} \nabla_{\mu \tau \alpha}R_{\nu \sigma \rho \beta} 
-  \tfrac{42}{25} \nabla^{\sigma}\hat\varphi_{\sigma}{}^{\tau \alpha} \nabla_{\mu}{}^{\beta}{}_{\beta}R_{\nu \tau \rho \alpha} 
+ \tfrac{98}{25} \nabla^{\sigma}\hat\varphi_{\mu}{}^{\tau \alpha} \nabla_{\nu}{}^{\beta}{}_{\beta}R_{\rho \tau \sigma \alpha} 
\nonumber\\ &
-  \tfrac{112}{25} \nabla_{\mu}\hat\varphi^{\sigma \tau \alpha} \nabla_{\sigma}{}^{\beta}{}_{\beta}R_{\nu \tau \rho \alpha} 
+ \tfrac{77}{25} \nabla^{\sigma}\hat\varphi_{\mu}{}^{\tau \alpha} \nabla_{\sigma}{}^{\beta}{}_{\beta}R_{\nu \tau \rho \alpha} 
+ \tfrac{42}{5} \nabla_{\sigma}R_{\nu \alpha \rho \beta} \nabla^{\sigma \tau}{}_{\tau}\hat\varphi_{\mu}{}^{\alpha \beta} 
\nonumber\\ &
+ \tfrac{56}{5} \nabla_{\alpha}R_{\nu \sigma \rho \beta} \nabla^{\sigma \tau}{}_{\tau}\hat\varphi_{\mu}{}^{\alpha \beta} 
-  \tfrac{98}{5} \nabla_{\alpha}R_{\rho \sigma \tau \beta} \nabla^{\sigma \tau \alpha}\hat\varphi_{\mu \nu}{}^{\beta} 
+ \tfrac{399}{50} \nabla_{\nu}R_{\rho \sigma \tau \beta} \nabla^{\sigma \tau \alpha}\hat\varphi_{\mu \alpha}{}^{\beta} 
\nonumber\\ &
+ \tfrac{721}{50} \nabla_{\tau}R_{\nu \sigma \rho \beta} \nabla^{\sigma \tau \alpha}\hat\varphi_{\mu \alpha}{}^{\beta} 
-  \tfrac{161}{50} \nabla_{\beta}R_{\nu \sigma \rho \tau} \nabla^{\sigma \tau \alpha}\hat\varphi_{\mu \alpha}{}^{\beta} 
- 7 \nabla_{\mu}R_{\nu \sigma \rho \beta} \nabla^{\sigma \tau \alpha}\hat\varphi_{\tau \alpha}{}^{\beta} 
\nonumber\\ &
+ \tfrac{154}{25} \nabla^{\sigma}\hat\varphi_{\mu}{}^{\tau \alpha} \nabla_{\tau}{}^{\beta}{}_{\beta}R_{\nu \sigma \rho \alpha} 
-  \tfrac{36}{5} \hat\varphi^{\sigma \tau \alpha} \nabla_{\mu \sigma}{}^{\beta}{}_{\beta}R_{\nu \tau \rho \alpha} 
-  \tfrac{42}{5} R_{\mu}{}^{\sigma \tau \alpha} \nabla_{\nu \rho \tau}{}^{\beta}\hat\varphi_{\sigma \alpha \beta} 
\nonumber\\ &
+ \tfrac{84}{5} R_{\mu}{}^{\sigma \tau \alpha} \nabla_{\nu \sigma \tau}{}^{\beta}\hat\varphi_{\rho \alpha \beta} 
+ \tfrac{56}{5} R_{\mu}{}^{\sigma \tau \alpha} \nabla_{\nu \tau}{}^{\beta}{}_{\beta}\hat\varphi_{\rho \sigma \alpha} 
-  \tfrac{42}{5} R_{\mu}{}^{\sigma}{}_{\nu}{}^{\tau} \nabla_{\rho \sigma}{}^{\alpha \beta}\hat\varphi_{\tau \alpha \beta} 
\nonumber\\ &
-  \tfrac{42}{5} R_{\mu}{}^{\sigma}{}_{\nu}{}^{\tau} \nabla_{\rho}{}^{\alpha \beta}{}_{\beta}\hat\varphi_{\sigma \tau \alpha} 
-  \tfrac{98}{5} R_{\mu}{}^{\sigma \tau \alpha} \nabla_{\sigma \tau}{}^{\beta}{}_{\beta}\hat\varphi_{\nu \rho \alpha} 
+ \tfrac{56}{5} R_{\mu}{}^{\sigma}{}_{\nu}{}^{\tau} \nabla_{\sigma}{}^{\alpha \beta}{}_{\beta}\hat\varphi_{\rho \tau \alpha} 
\nonumber\\ &
+ \tfrac{21}{5} R_{\mu}{}^{\sigma}{}_{\nu}{}^{\tau} \nabla^{\alpha}{}_{\alpha}{}^{\beta}{}_{\beta}\hat\varphi_{\rho \sigma \tau} 
-  \tfrac{2}{5} \nabla_{\mu \nu \rho}{}^{\sigma \tau \alpha}\hat\varphi_{\sigma \tau \alpha} 
+ \tfrac{12}{5} \nabla_{\mu \nu}{}^{\sigma \tau \alpha}{}_{\alpha}\hat\varphi_{\rho \sigma \tau} 
\nonumber\\ &
- 3 \nabla_{\mu}{}^{\sigma \tau}{}_{\tau}{}^{\alpha}{}_{\alpha}\hat\varphi_{\nu \rho \sigma} 
+ \nabla^{\sigma}{}_{\sigma}{}^{\tau}{}_{\tau}{}^{\alpha}{}_{\alpha}\hat\varphi_{\mu \nu \rho}
+ \mathcal{O}(R^2) ,
\end{align}
where $\nabla_{\mu_1 \cdots \mu_n} = \nabla_{(\mu_1} \cdots \nabla_{\mu_n)}$ and
$\hat\varphi_{\mu \nu \rho} = \varphi_{\mu \nu \rho} - \tfrac{1}{2} g_{(\mu\nu} \varphi_{\rho)\sigma}{}^\sigma$.

\section{Conclusions}
\label{Sec:Conclusions}

In this paper we have studied conformal wave operators for HS fields on general
backgrounds. We have found a manifestly factorized form for them in (A)dS, and
for spin 2 on arbitrary Einstein backgrounds. The whole analysis has been
carried out in arbitrary dimensions. The main result of this paper is the
explicit form of the wave operator on (A)dS backgrounds, together with the
identification of the obstruction to factorization on more general backgrounds.

The results of this paper confirm previous conjecture about conformal HS wave
operators on (A)dS backgrounds \cite{Joung:2012qy,Tseytlin:2013jya}. On the
other hand the identification
of the obstruction to factorization for spin $s>2$ HS wave operators on more
general backgrounds lead us to reconsider modifications of this conjecture.
Specifically, the computation of the $c$-coefficient of the Weyl anomaly done in
\cite{Tseytlin:2013jya}, which assumes factorization on Ricci-flat backgrounds,
should be reconsidered.

We expect the variant of the Tractor formalism exploited in this paper to be a
key tool for further analysis of conformal HS theories on generic backgrounds.
We plan to come back to these issues in future publications. The full form
of the conformal wave operator on generic backgrounds is still missing, and so
far we have been able to fix it only up to linear order in the Riemann tensor
for spin 3.

Before concluding let us mention once again that the operator gauge invariance
condition turns to be very powerful to control conformal invariance in any
dimension. Therefore, we conjecture the existence of a solution to the latter
stronger operator condition on general backgrounds. This feature can be also
interpreted by saying that operator gauge invariance of the corresponding wave
operator is equivalent to its conformal invariance. Since in our setting we only
require linear Weyl symmetry on top of gauge symmetry, this observation shares
possible similarities with analogous statements in the context of CFT (see e.g.
\cite{Dymarsky:2013pqa,Bzowski:2014qja,Dymarsky:2014zja}).

It will also be interesting to address questions about interactions and gauge
algebra deformations with the variant of the tractor calculus introduced here.
We leave this as well as other interesting questions related to conformal HS
fields for future research.

\section*{Note added} 

During the final stages of preparation of the present article the paper
\cite{Metsaev:2014iwa} by R.~Metsaev appeared. Although using different
techniques, it contains some results that are in overlap with the results
presented in \autoref{Sec:Zeroth_Order}. While we use an explicitly higher
derivative formalism, \cite{Metsaev:2014iwa} exploits an ordinary derivative
formulation by introducing auxiliary fields. The results of
\cite{Metsaev:2014iwa} are equivalent to the factorization of the conformal
operator in (A)dS background that we recover in a different way.

\section*{Acknowledgments}

We are indebted to E.~Skvortsov and S.~Theisen for useful discussions and
comments, and for having brought this problem to our attention. We are also
grateful to R.~Manvelyan and A.~Sagnotti for useful discussions and comments on
the manuscript. We performed various computations with the \emph{xAct} collection
of Mathematica packages \cite{MartinGarcia:2002xa}, and in particular with
\emph{xTras} \cite{Nutma:2013zea}.

\appendix

\section{Notation and conventions}
\label{Appendix:Relations}

In this appendix we give a brief introduction to the techniques and conventions
we used to deal with conformal HS fields.

We mainly rely on an operator formalism where index contraction and
symmetrization of indices are realized in terms of auxiliary variables. This
allows us to translate tensor operations in terms of operator calculus,
resulting in simplified manipulations (see e.g. \cite{Taronna:2012gb} for
further details).

After replacing symmetric tensors by polynomials in the auxiliary variable $u^a$
as in equation \eqref{eq:genfunc}, it is possible to define the action of the
covariant derivative as a differential operator on both $x$ and $u$:
\begin{subequations}
\begin{align}
	\tilde\nabla_\mu  \rightarrow \nabla_{\mu}&=\tilde\nabla_{\mu}-\tfrac12\omega_{\mu\ b}^{\ a}L_a^{\ b} =
		\tilde\nabla_{\mu}-\omega_{\mu\ b}^{\ a}u^b\partial_{u^a},\\
	[\nabla_\mu,\nabla_\nu]& =
		\Lambda(u_\mu\partial_{u_\nu}-u_\nu\partial_{u_\mu})
		+R^{\Lambda}_{\mu\nu\rho\sigma}(x)u^\rho\partial_{u_\sigma},
\end{align}
\end{subequations}
where above and henceforth commutator equations will be assumed to hold on scalar functions of
$u$ with no naked tensorial index.
Here $\tilde\nabla_\mu$ is the standard covariant derivative acting on naked 
tensorial indices, $\omega$ is the spin-connection and $L_a^{\ b}$ are the Lorentz generators.
We have expressed the latter in terms of differential operators upon introducing
the derivative $\partial_{u^a}$, which is defined by:
\begin{align}
	&\partial_{u^a} u^b=\delta_a^b.& &L^a_{\ b}=u^a\partial_{u^b}-u^b\partial_{u^a}.
\end{align}
We have also expressed the commutator of covariant derivatives in terms of 
$R^\Lambda_{\mu\nu\rho\sigma}$. This is simply the Riemann tensor minus its 
constant trace part:
\begin{equation}
	R^{\Lambda}_{\mu\nu\rho\sigma} = 
		R_{\mu\nu\rho\sigma} 
		- \Lambda (g_{\mu\rho}g_{\nu\sigma} - g_{\nu\rho}g_{\mu\sigma}),
\end{equation}
This conveniently parametrizes the difference between constant curvature metrics
and more general ones.

In what follows we shall work only with the contracted auxiliary variable
$u^\mu=e^{\ \mu}_{a}(x)u^{a}$ and the associated derivative $\partial_{u^\mu} = 
e^{a}_{\ \mu}(x)\partial_{u^{a}}$. The latter commutes with the covariant 
derivative on generic backgrounds as a consequence of the vielbein postulate:
\begin{align}
	[\nabla_\mu,u^\nu] & =0, &
	[\partial_{u^\mu},\nabla_\nu] & = 0.
\end{align}
The operators box, symmetrized gradient, divergence, trace, symmetrized metric,
and spin can then be represented respectively by the following operators:
\begin{align}
	\textrm{box:  } 			& \dal, & 
	\textrm{divergence:  } 		& \nabla\cdot\partial_u,&
	\textrm{sym.~metric:  } 	& u^2, \nonumber \\
	\textrm{sym.~gradient:  } 	& u\cdot\nabla,	&
	\textrm{trace:  } 			& \partial_u^2,& 
	\textrm{spin:  } 			& u\cdot\partial_u .
\end{align}
They satisfy the following operator algebra:
\begin{subequations}
\begin{align}
	[\dal,u\cdot\nabla]&=\Lambda \left[u\cdot\nabla(2u\cdot\partial_u+d-1)-2u^2\nabla\cdot\partial_u\right]\\
					   &+2R^{\Lambda}_{\mu\nu\rho\sigma}\nabla^\mu u^\nu u^\rho\partial_{u_\sigma}-(\nabla_{\sigma} R^{\Lambda}_{\nu\rho} - \nabla_\rho R^{\Lambda}_{\nu\sigma})u^\nu u^\rho\partial_{u^\sigma}+R^\Lambda_{\nu\rho}u^\nu\nabla^\rho,\nonumber\\
	[\nabla\cdot\partial_u,\dal]&=\Lambda\left[(2u\cdot\partial_u+d-1)\nabla\cdot\partial_u-2u\cdot \nabla \partial_u^2\right]\\
								&-2R^{\Lambda}_{\mu\nu\rho\sigma}\nabla^{\mu}u^\rho \partial_{u^\sigma}\partial_{u^\nu} +R^{\Lambda}_{\mu\nu}\nabla^\mu\partial_{u^\sigma}+(\nabla^\mu R^{\Lambda}_{\mu\sigma})\partial_{u_\sigma} \nonumber\\
								&- (\nabla_\sigma R^{\Lambda}_{\nu\rho}-\nabla_\rho R^{\Lambda}_{\nu\sigma})u^\rho \partial_{u_\sigma}\partial_{u_\nu}\nonumber\\
	[\nabla\cdot\partial_u,u\cdot\nabla]&=\dal+\Lambda\left[u\cdot\partial_u(u\cdot\partial_u+d-2)-u^2\partial_u^2\right]\nonumber\\&+R^\Lambda_{\mu\nu\rho\sigma}u^\nu u^\rho \partial_{u_\mu}\partial_{u_\sigma}+R^\Lambda_{\mu\nu}u^\mu \partial_{u_\nu},\\
	[\nabla\cdot\partial_u,u^2]&=2u\cdot\nabla,\\
	[\partial_u^2,u\cdot\nabla]&=2\nabla\cdot\partial_u,\\
	[\partial_u^2,u^2]&=2(d+2u\cdot\partial_u).
\end{align}
\end{subequations}
On Einstein backgrounds these commutation relations simplify due to the identity
$R^\Lambda_{\mu\nu\rho\sigma}=W_{\mu\nu\rho\sigma}$, where
$W_{\mu\nu\rho\sigma}$ is the Weyl tensor.  The main difficulty is however the
fact that the  operator algebra does not close and requires the inclusion of
Riemann tensors and their derivatives of arbitrary order. The algebra closes
only if one restricts it to its spin $s$ sector.

In the case of conformal higher spin fields one needs to work with fields defined
on equivalence classes,
\begin{equation}
	\varphi_{\mu_1\cdots\mu_s} \sim 
		\varphi_{\mu_1\cdots\mu_s}+g_{(\mu_1\mu_2}\alpha_{\mu_3\cdots\mu_s)},
\end{equation}
or in terms of the auxiliary variables:
\begin{equation}
\label{eq:confeq}
	\varphi\sim \varphi +u^2\alpha.
\end{equation}
In order to work on such equivalence classes it is quite useful to exploit a
variant of the Tractor calculus (see e.g. \cite{Joung:2013doa} and references
therein) in which one replaces ordinary derivative operators $\partial_{u}$ with
Thomas-D derivatives:
\begin{equation}
	\hat\partial_{u^\mu}=\partial_{u^\mu}-\frac{1}{h}u_\mu\partial_u^2.
\end{equation}
Here we have defined $h$ as
\begin{equation}
	h=d-2+2u\cdot\partial_u.
\end{equation}
Thomas-D derivatives have the useful property to be automatically defined on the
above equivalence classes, since
\begin{equation}
	\hat\partial_{u^\mu}u^2 = 
		u^2\left(\partial_{u^\mu}-\tfrac{1}{h-4}u_\mu\partial_u^2\right)\sim0.
\end{equation}
In this way the operator algebra simplifies since we can consistently set 
$u^2\sim 0$, and we end up with only four operators: $\dal$, $u\cdot\nabla$, 
$\nabla\cdot \hat\partial_u$, and $u\cdot\hat\partial_u$. Notice that
\begin{equation}
	\hat\partial_{u}^2=u^2 (\partial_u^2)^2\sim 0.
\end{equation}
Further using the commutation relation
\begin{equation}
	[\hat\partial_{u^\mu},u^\nu] = 
		g_{\mu\nu}-\tfrac{2}{h}u_\mu\hat\partial_{u^\mu},
\end{equation}
we end up with the following operator algebra:
\begin{subequations}\label{eq:operatoralgebra2}
\begin{align}
[\nabla_\mu,\nabla_\nu]&=\Lambda(u_\mu\hat\partial_{u_\nu}-u_\nu\hat\partial_{u_\mu})+R^{\Lambda}_{\mu\nu\rho\sigma}(x)u^\rho\hat\partial_{u_\sigma},\\
[\dal,u\cdot\nabla]	&=\Lambda u\cdot\nabla(2u\cdot\hat\partial_u+d-1)\\
					&-2R^{\Lambda}_{\mu\nu\rho\sigma}u^\mu\nabla^\nu  u^\rho\hat\partial_{u_\sigma}-
					u^\nu u^\rho(\nabla\cdot\hat\partial_u R^{\Lambda}_{\nu\rho}) + (u\cdot\nabla R^{\Lambda}_{\nu\sigma})u^\nu\hat\partial_{u^\sigma}+R^\Lambda_{\mu\nu}u^\mu\nabla^\nu,\nonumber\\
[\nabla\cdot\hat\partial_u,\dal]&=\Lambda(2u\cdot\hat\partial_u+d-1)\nabla\cdot\hat\partial_u\\
					&-2R^{\Lambda}_{\mu\nu\rho\sigma}\nabla^{\mu}u^\rho \hat\partial_{u^\nu}\hat\partial_{u^\sigma} +R^{\Lambda}_{\mu\nu}\nabla^\mu\hat\partial_{u^\nu}+(\nabla^\mu R^{\Lambda}_{\mu\sigma})\hat\partial_{u_\sigma} \nonumber\\
					&+ u^\rho \hat\partial_{u_\nu}(\nabla\cdot\hat\partial_u R^{\Lambda}_{\nu\rho})-
					(u\cdot\nabla R^{\Lambda}_{\nu\sigma})\hat\partial_{u_\nu}\hat\partial_{u_\sigma}\nonumber,\\
[\nabla\cdot\hat\partial_u,u\cdot\nabla]&=\dal-\tfrac{2}{h}u\cdot\nabla\nabla\cdot\hat\partial_u+\Lambda u\cdot\hat\partial_u(u\cdot\hat\partial_u+d-2)\\
&-R^\Lambda_{\mu\nu\rho\sigma}u^\mu u^\rho \hat\partial_{u_\nu}\hat\partial_{u_\sigma}+R^\Lambda_{\mu\nu}u^\mu \hat\partial_{u_\nu}\nonumber .
\end{align}
\end{subequations}
This operator algebra is defined on equivalence classes \eqref{eq:confeq},
and again closes only if one also includes derivatives of the Riemann tensor 
and their commutators recursively.

\section{Spin s wave operator in standard tensor notation}

It is not too difficult to present the generic recursive structure of the two derivative operators entering the (A)dS solution in terms of standard tensor notation.
One can then define the following recursion relation
\begin{multline}
\varphi^{(i-1)}_{\mu(s)}=\mathcal{P}_{\mu(s)}^{\quad\ \nu(s)}\Bigg\{\Big[\dal-\Lambda[(i-s+1)(i-s-d+2)-s\Big]\varphi^{(i)}_{\nu(s)}\\-\tfrac{d-4+2s}{i(d-3-i+2s)}\left[s\nabla_{\nu}\nabla^\alpha\varphi^{(i)}_{\alpha\nu(s-1)}+\tfrac{s(s-1)}{d-4+2s}\nabla_{\nu}\nabla_{\nu}\varphi^{(i)\alpha}_{\quad\ \alpha\nu(s-2)}\right]\Bigg\},
\end{multline}
where eliminating the auxiliary variable acting with the operator $(\hat\partial_{u_\mu})^s$, we are left with the spin s traceless projector $\mathcal{P}_{\mu(s)}^{\quad\ \nu(s)}$.
Above, we have conveniently defined new fields $\varphi^{(i-1)}_{\mu(s)}$ with $\varphi_{\mu(s)}^{(0)}=\mathcal{O}_{\mu(s)}^{(s)}$ and $\varphi_{\mu(s)}^{\left(\tfrac{d}{2}-2+s\right)}=\varphi_{\mu(s)}$ of weight shifting by two units at each step. One then ends up with the conformal operator written in standard tensor notation upon substituting the corresponding fields above till expressing $\varphi_{\mu(s)}^{(0)}$ in terms of $\varphi_{\mu(s)}$.

\section{Wave operator in non factorized form}
\label{Appendix:ConformalWaveNonF}

In this appendix we will rewrite the factorized wave operator for a conformal
spin $s$ field on (A)dS backgrounds in a more standard form from which one can 
read off the analogue of the de Donder tensor for conformal higher spins.

We start by writing an Ansatz of the type:
\begin{align}\label{eq:Ansatz}
{\mathcal O}^{(s)}&=\sum_{i=0}^{s+\tfrac{d}{2}-2}\gamma_i (u\cdot\nabla)^i{\mathcal B}_{s+\tfrac{d}{2}-2-i}(\nabla\cdot\hat\partial_u)^i\\&=\sum_{i=0}^{s+\tfrac{d}{2}-2}\gamma_i (u\cdot\nabla)^i\left[\prod_{j=1}^{s+\tfrac{d}{2}-2-i}\nonumber
(\dal+\beta_{i,j}\Lambda)\right](\nabla\cdot\hat\partial_u)^i.
\end{align}
A useful trick is then to parameterize the gauge variation of a divergence as:
\begin{equation}
(\nabla\cdot\hat\partial_u)^n u\cdot\nabla = 
	\left[
	 	a_n\dal+b_n u\cdot\nabla\nabla\cdot\hat\partial_u+\Lambda c_n
	 \right](\nabla\cdot\hat\partial_u)^{n-1},
\end{equation}
where the coefficients satisfy the following recursion relations
\begin{subequations}
\begin{align}
	a_n&=a_{n-1}+b_{n-1},\\
	b_n&=b_{n-1}\left(1-\tfrac{2}{d-2+2(s-n)}\right),\\
	c_n&=c_{n-1}+b_{n-1}(s-n)(s-n+d-2)+a_{n-1}(2(s-n)+d-1),
\end{align}
\end{subequations}
with
\begin{equation}
	a_1=1,\qquad b_1=-\frac{2}{d-4+2s},\qquad c_1=(s-1)(s+d-3),
\end{equation}
and hence
\begin{subequations}
\begin{align}
	a_n&=1-\frac{n(n-1)}{d-4+2s}+\frac{(n-1)(n-2)}{d-6+2s},\\
	b_n&=-\frac{\tfrac{d}{2}-2+s-n}{(\tfrac{d}{2}-2+s)(\tfrac{d}{2}-3+s)},
\end{align}
\end{subequations}
while we do not present the solution for $c_n$ for brevity.
One can now compute the gauge variation of the operator ${\mathcal E}_i=\gamma_i (u\cdot\nabla)^i{\mathcal B}_{s+\tfrac{d}{2}-2-i}(\nabla\cdot\hat\partial_u)^i$:
\begin{align}
{\mathcal E}_i u\cdot\nabla&=\gamma_i a_i(u\cdot\nabla)^{i}\left[\prod_{j=1}^{s+\tfrac{d}{2}-2-i}
(\dal+\Lambda\beta_{i,j})\right](\dal+\Lambda \tfrac{c_i}{a_i})(\nabla\cdot\hat\partial_{u})^{i-1}\\
&+\gamma_i b_i (u\cdot\nabla)^{i+1}\left[\prod_{j=1}^{s+\tfrac{d}{2}-2-i}
\left[\dal+\Lambda\left(\beta_{i,j}+2(s-i)+d-3\right)\right]\right](\nabla\cdot\hat\partial_{u})^{i}.\nonumber
\end{align}
Therefore, by requiring that the terms proportional to $(u\cdot\nabla)^{i+1}$ in
the variation of ${\mathcal E}_i$ cancel the terms proportional to
$(u\cdot\nabla)^{i+1}$ in the variation of ${\mathcal E}_{i+1}$ one gets the
following conditions for the free coefficients $\gamma_i$ and $\beta_{i,j}$:
\begin{subequations}
\begin{align}
	\gamma_{i+1}&=-\frac{b_i}{a_{i+1}}\gamma_i,\\
	\beta_{i,1}&=\frac{c_{i+1}}{a_{i+1}}-2(s-i)-d+3,\\
	\beta_{i,j}&=\beta_{i+1,j-1}-2(s-i)-d+3.
\end{align}
\end{subequations}
The conditions can be solved to give
\begin{subequations}
\begin{align}
	\gamma_i&=(-1)^i\frac{\prod_{n=0}^{i-1}b_n}{\prod_{n=1}^{i}a_n},\qquad \gamma_0=1,\\
	\beta_{i,1}&=\frac{c_{i+1}}{a_{i+1}}-2(s-i)-d+3,\\
	\beta_{i,j}&=\beta_{i+j-1,1}-(j-1)[2(s-i)+d-j-1].
\end{align}
\end{subequations}
After plugging in the solution for the coefficients $a_i$, $b_i$ and $c_i$ we
then get
\begin{equation}
	\beta_{i,j}=(i+j+1-s)(i+j-s-d+2)-(j-1)[2(s-i)+d-j-1]-s.
\end{equation}
As before, this matches all partially massless points in $d=4$, and also some massive points
in higher dimensions.

The generalized de Donder tensor can be easily extracted from equation 
\eqref{eq:Ansatz}:
\begin{equation}
\label{eq:dedondertensor}
	{\mathcal D}_{\left(\tfrac{d}{2}-3+s\right)} = \sum_{i=1}^{\tfrac{d}{2}-2+s} 
			(u\cdot\nabla)^{i-1} \mathcal{B}_{s+\tfrac{d}{2}-2-i} (\nabla\cdot\hat\partial_u)^i \varphi_{(s)} .
\end{equation}
This tensor has one derivative less than the full equation of motion.
From the gauge invariance condition one can easily extract its gauge variation:
\begin{equation}
	\delta_\xi \mathcal{D}_{\left(\tfrac{d}{2}-3+s\right)} 
		= - \prod_{j=1}^{\tfrac{d}{2}-2+s} \Bigl( \dal + (\beta_{0,j} + 2 s +d - 3) \Lambda\Bigr) \varepsilon_{(s-1)} .
\end{equation}
The right-hand-side can be viewed as a second order equation on an effective
gauge parameter that is of order $2s+d-6$. This linear second order diagonal
equation can be solved throughout spacetime \cite{Hawking1973} in order to 
set ${\mathcal D}_{\left(\tfrac{d}{2}-3+s\right)}$ to zero. In this partial 
gauge, the equation of motion becomes \eqref{eq:eomTT}.

\section{Examples in various dimensions}\label{Appendix:Examples}

In this appendix we list some known non-linear conformal actions, and confirm
that their equations of motion reduce to \eqref{eq:Conf_Operator} upon
linearization on (A)dS spaces.

\subsection{Spin 1 data}

\paragraph{d=2} 

The 2 dimensional case is trivial since the spin $1$ conformal field does not
propagate and indeed the number of derivatives compatible with conformal
symmetry is 0.

\paragraph{d=4} 

In four dimensions the Maxwell's theory is conformally invariant, and its equation of
motion is precisely \eqref{eq:Ansatz} for $s=1$ and $d=4$.

\paragraph{d=6} 

In six dimensions there are a number conformal invariants quadratic in $A = \varphi_{(1)}$.
Yet there is only one that is gauge invariant, not a total derivative, and
non-zero on AdS backgrounds. It reads
\begin{equation}
	I = F^{\mu\nu} 
	\Bigl( 
		( \dal - \tfrac{1}{2} R ) \delta^\rho_\mu \delta^\sigma_\nu
		+ R_\mu{}^\rho \delta_\nu^\sigma
		+ C_{\mu\nu}{}^{\rho\sigma}
	\Bigr) F_{\rho\sigma} + \nabla_\mu J^\mu ,
\end{equation}
with $F_{\mu\nu} = \nabla_{[\mu} A_{\nu]}$. The Weyl tensor could have been omitted,
as $F \cdot C \cdot F$ is conformally invariant on its own. However, including it
reproduces Branson's $D_{4,1}$ conformal operator \cite{branson1985} acting
on $A_\mu$ as the equation of motion:
\begin{equation}
	\nabla^\nu \bigl(
		\nabla_{[\mu} \nabla^\rho F_{\nu]\rho} + S F_{\mu\nu} - 4 S^\rho{}_{[\mu} F_{\nu]\rho}
 	\bigr) = 0 ,
\end{equation}
where $S_{\mu\nu}$ is the Schouten tensor and $S$ is its trace.
Upon linearizing these equations of motion on (A)dS we find \eqref{eq:Conf_Operator} or \eqref{eq:Ansatz}
for $s=1$ and $d=6$ in agreement with the solution to the operator gauge invariance condition.

\subsection{Spin 2 data}
\label{sec:spin2data}

\paragraph{d=2} 

Two-dimensional conformal gravity is just Einstein gravity, whose linearized
equation of motion on (A)dS can be precisely recast in the form \eqref{eq:Conf_Operator} or \eqref{eq:Ansatz}
for $s=2$ and $d=2$.

\paragraph{d=4} 

The action for four dimensional conformal gravity is
\begin{equation}
	S = \int d^4 x \sqrt{-g} C_{\mu\nu\rho\sigma} C^{\mu\nu\rho\sigma} ,
\end{equation}
whose linearized equation of motion is exactly \eqref{eq:Conf_Operator} or \eqref{eq:Ansatz} for $s=2$ and $d=4$.

\paragraph{d=6} 

In six dimensions there are three conformal invariants for gravity, namely
\cite{Bastianelli:2000rs,Metsaev:2010kp,Oliva:2010zd}
\begin{subequations}
\begin{align}
	I_1 & = C_{\mu\rho\sigma\nu} C^{\mu\alpha\beta\nu} C_{\alpha}{}^{\rho\sigma}{}_\beta , \\
	I_2 & = C_{\mu\nu\rho\sigma} C^{\rho\sigma\alpha\beta} C_{\alpha\beta}{}^{\mu\nu} , \\
	I_3 & = C_{\mu\rho\sigma\lambda} \Bigl(
		\delta^\mu_\nu \dal + 4 R^\mu{}_\nu - \tfrac{6}{5} \delta^\mu_\nu R 
	\Bigr) C^{\nu\rho\sigma\lambda} + \nabla_\mu J^\mu
\end{align}
\end{subequations}
with $\nabla_\mu J^\mu$ a total derivative which can be found in \cite{Bastianelli:2000rs}.
Because the Weyl tensor vanishes on AdS backgrounds, only the third invariant gives a
non-zero quadratic perturbation on AdS. Upon computing its equations of motion, we find
\eqref{eq:Conf_Operator} or \eqref{eq:Ansatz}
for $s=2$ and $d=6$, again in agreement with the general result obtained above enforcing the stronger operator gauge invariance condition.

\begingroup
\setlength{\emergencystretch}{8em}
\printbibliography
\endgroup

\end{document}